\documentclass[12pt]{article}
\usepackage{graphics}
\newcommand{\la}[1]{\label{#1}}
\def\nn{\nonumber}

\newcommand{\bfn}{{{\bf \hat n}}}

\newcommand{\bfF}{{\bf F}}
\newcommand{\vecn}{{\bf \hat n}}
\newcommand{\vece}{{\bf \hat e}}

\newcommand{\be}{\begin{equation}}
\newcommand{\ee}{\end{equation}}
\newcommand{\ba}{\begin{eqnarray}}
\newcommand{\ea}{\end{eqnarray}}
\newcommand{\bastar}{\begin{eqnarray*}}
\newcommand{\eastar}{\end{eqnarray*}}

\title{Three Dimensional Gravity From \\ \vskip 0.4cm 
   SU(2) Yang-Mills Theory in Two Dimensions }
\author{ A.J.\ Niemi\thanks{Antti.Niemi@teorfys.uu.se}\hspace{1cm}
         \\ \\
         {\it Department of Theoretical Physics}\\
         {\it Uppsala University}\\
         {\it Box 803, SE-751 08 Uppsala, Sweden}}
\begin{document}

\maketitle

\begin{abstract}

\noindent
We argue that two dimensional classical SU(2) Yang-Mills theory 
describes the embedding of Riemann surfaces in three 
dimensional curved manifolds. Specifically, the 
Yang-Mills field strength tensor computes
the Riemannian curvature tensor of the ambient space in a 
thin neighborhood of the surface. In this sense the
two dimensional gauge theory then serves as a source of three
dimensional gravity. In particular, if the three 
dimensional manifold is flat it corresponds to the 
vacuum of the Yang-Mills theory.
This implies that all solutions to the original
Gauss-Codazzi surface equations 
determine two dimensional integrable models with
a SU(2) Lax pair. Furthermore, the three dimensional
SU(2) Chern-Simons theory describes the Hamiltonian dynamics 
of two dimensional Riemann surfaces 
in a four dimensional flat space-time.
\end{abstract}
\vfill\eject

{
\baselineskip 0.6cm

\vskip 1.5cm
\noindent
{\bf \Large 1. Introduction}
\vskip 1.0cm
The isometric embedding of a two dimensional Riemann surface in a three
dimensional ambient space is a classic problem in differential
geometry \cite{spivak}. The embedding involves the first and
second fundamental forms of the surface, engaging 
its intrinsic metric and extrinsic curvature. 
Together the two fundamental forms determine the metric of 
the three dimensional ambient space, in a vanishingly thin 
neighborhood around the two dimensional surface. 
If the three dimensional space is flat $R^3$  
the two dimensional metric and curvature are subject to the 
original version of the 
Gauss-Codazzi equations. But when these equations 
are invalid the three dimensional curvature is  
non-trivial, at least in an immediate vicinity of the 
surface. In this sense the two dimensional surface
is then a source of gravity in the three dimensional 
ambient space. 

In the present article we shall assert that similarly
the two dimensional
SU(2) Yang-Mills field can be viewed as an origin of 
three dimensional Riemannian
curvature. Indeed, there are suggestions that something
like this could occur. For example, it is well known that 
solutions to the sine-Gordon equation 
\ba
\omega_{tt} - \omega_{xx} \ +  \ \sin \omega \ = \ 0 
\la{sg}
\ea
describe constant negative Gaussian curvature surfaces in $R^3$,
up to rigid Euclidean motions.
But the sine-Gordon equation also emerges as the zero curvature 
condition for the SU(2) Yang-Mills field strength tensor \cite{fadbook}
(in the sequel $i,j,k,... = 1,2$ and 
$\alpha,\beta,\gamma,... = 1,2,3$)
\ba
F^\alpha_{ij} \ = 
\ \partial_i A^\alpha_j - \partial_j A^\alpha_i
+ \epsilon^{\alpha\beta\gamma} A^\beta_i A^\gamma_j  \ = \ 0
\la{zeroF}
\ea
when we decompose the connection $A^\alpha_i$ according 
to \cite{fadbook}
\ba
A_1^\alpha \tau^\alpha \ = \ \frac{\omega_x}{2} \tau^3
\ + \ k_2 \sin \frac{\omega}{2} \ \tau^1 \ - \ k_1
\cos \frac{\omega}{2} \ \tau^2 \nn
\ea
\ba
A^\alpha_2 \tau^\alpha \ = \ \frac{\omega_t}{2} \tau^3 \ + \
k_1 \sin \frac{\omega}{2} \ \tau^1 \ - \ k_2 \cos 
\frac{\omega}{2} \ \tau^2
\la{sgdec}
\ea
where $k_1^2 - k_2^2 = 1$ and $k_i$ involve the
spectral parameter, real for a SU(2) 
connection. Similar relations between the
zero curvature condition (\ref{zeroF}) in the 
two dimensional SU(2) Yang-Mills theory 
and the embedding of a Riemann surface in flat $R^3$ 
have been established for a number of 
additional integrable models \cite{lund}.

In the present article we shall argue that for SU(2)
the condition (\ref{zeroF}) always
describes the embedding of a Riemann
surface in $R^3$. For this we shall consider a scrupulous
decomposition of $A^\alpha_i$ which reveals that the condition
(\ref{zeroF}) coincides with the Gauss-Godazzi 
surface equations that govern the isometric 
embedding of a Riemann surface in flat $R^3$, up to rigid
rotations and translations. Consequently any 
integrable model with a SU(2) Lax pair always admits an interpretation 
in terms of a Riemann surface which is isometrically 
embedded in the flat three dimensional Euclidean space.

Furthermore, we shall employ our decomposition of 
the two dimensional $A^\alpha_i$
to establish that whenever the
condition (\ref{zeroF}) fails, the Yang-Mills
field strength tensor $F^\alpha_{ij}$ leads to a non-trivial 
three dimensional Riemannian curvature tensor in a thin neighborhood 
of the two dimensional hypersurface. This implies that the two 
dimensional Yang-Mills theory induces gravity in three dimensions. 
The tentative consistency of this proposal can be
verified by comparing the number of 
field degrees of freedom: The two dimensional
SU(2) gauge field $A^\alpha_i$ has six components, which are
subject to three Gauss law constraints. Similarly, the three dimensional
metric $G_{\mu\nu}$ ($\mu,\nu = 1,2,3$) 
has six independent matrix elements, and these are  
amenable to diffeomorphisms which involve three
field degrees of freedom. Thus both two dimensional $A^\alpha_i$
and three dimensional $G_{\mu\nu}$ carry the same number of
field degrees of freedom, with an equal number of gauge degrees of freedom. 

We shall now proceed to establish the relationship between
the two dimensional gauge theory and three dimensional
gravity beyond such a simple counting of field degrees freedom. 
We show that the two dimensional 
Yang-Mills field strength tensor actually computes the
three dimensional Riemannian curvature tensor in the immediate 
vicinity of the two dimensional hypersurface.

\vskip 1.5cm
\noindent
{\bf \Large 2. Decomposition}
\vskip 1.0cm

The decomposition of vectors and tensors in terms of 
their irreducible components is a common problem
in Physics. For example, it is widely employed in fluid dynamics 
where the velocity three vector decomposes into its
gradient and vorticity components. In classical
electrodynamics the four dimensional
Maxwellian field strength tensor
$F_{\mu\nu}$ becomes similarly dissected into its electric 
and magnetic components. In the context of
integrable models, the Lax pair representation leads
to decompositions (\ref{sgdec}) of the 
two dimensional non-abelian gauge field in terms of 
variables that describe the integrable model. 
Finally, the isometric embedding of a 
two dimensional surface (in general 
$d$-dimensional hypersurface) 
with local coordinates $y^i$
in a three dimensional (in general $d+n$ dimensional)
ambient space with local coordinates $x^\mu$ and metric 
$G_{\mu\nu}$, involves the decomposition of 
the induced metric 
\ba
ds^2 \ = \ g_{ij}dy^i dy^j \ = \ G_{\mu\nu} 
\partial_i
x^\mu \partial_j x^\nu dy^i dy^j \nn
\ea
This embedding also entails the decomposition 
(Gauss equation)
\ba
\partial_{ij} x^\mu \ + \ {\hat \Gamma}^\mu_{\nu\rho}
\partial_i x^\nu \partial_j x^\rho \ = \ 
\Gamma^{k}_{ij} \partial_k x^\mu \ + \
Q_{ij} N^\mu 
\la{gwe}
\ea
Here ${\hat\Gamma}^\mu_{\nu\rho}$ is the metric connection
in the three dimensional 
ambient space, $\Gamma^k_{ij}$ 
is the (induced) metric connection on 
the two dimensional hypersurface and $Q_{ij}$ 
is its extrinsic curvature tensor, and $N^\mu$
is the three dimensional unit normal of the hypersurface.

Here we inspect how (\ref{gwe}) relates
to the following decomposition of the two dimensional SU(2) Yang-Mills
gauge field $A^\alpha_i$, introduced originally in \cite{fadprl}
\be
A^\alpha_i \ = \ C_i n^\alpha + \epsilon^{\alpha\beta\gamma}
\partial_i n^\beta n^\gamma +
\rho \partial_i n^\alpha + \sigma \epsilon^{\alpha\beta\gamma} 
\partial_i n^\beta n^\gamma
\la{dec1}
\ee
with $\rho,\sigma$ scalar fields.
Notice that we separate the second and 
fourth term on the {\it r.h.s}. and the reason for this
becomes evident as we proceed.  

We shall argue that in two dimensions (\ref{dec1}) is 
a complete decomposition of the full SU(2) Yang-Mills gauge field
into its irreducible components. 

Furthermore, we shall argue 
that (\ref{dec1}) admits a differential geometric interpretation
in terms of the quantities on the {\it r.h.s.} of (\ref{gwe}), describing
the isometric embedding of a Riemann surface in a three dimensional
ambient space. Specifically, the vector $C_i$ 
relates to the (induced) metric connection $\Gamma^k_{ij}$ on the 
two dimensional hypersurface, and $\rho$ and $\sigma$ 
relate to the two eigenvalues of its (symmetric) extrinsic 
curvature tensor $Q_{ij}$, and
the (three component) unit vector $n^\alpha$ maps to 
the (three component) unit normal $N^\mu$ by 
\[
n^\alpha \ = \ {e^\alpha}_\mu N^\mu
\]
with ${e^\alpha}_\mu$ a (flat) dreibein that relates
the two unit vectors; these vectors both 
reside in a flat $R^3$, the vector $n^\alpha$ is in the tangent bundle of 
the gauge group SU(2) while $N^\mu$ is a vector field in the 
ambient $R^3$, the normal map of the two dimensional hypersurface. 
These two spaces become identified by ${e^\alpha}_\mu$ which  
is obviously an element of SO(3).  

Finally, we shall argue that when we 
substitute the decomposition (\ref{dec1}) in the
Yang-Mills field strength tensor $F^\alpha_{ij}$ it produces 
the Riemann curvature tensor of the ambient space,
when evaluated in the immediate vicinity of the surface.

The decomposition (\ref{dec1}) was introduced and
inspected in \cite{fadprl},
in connection of four dimensional SU(2) Yang-Mills theory 
where it is known to be incomplete \cite{fadde2}. 
But we now argue that in two dimensions the 
decomposition (\ref{dec1}) is complete, describing 
the six independent components of a generic
two dimensional SU(2) gauge field $A^\alpha_i$. 

Indeed, when $D=2$ the vector field $C_i$ has two components. 
Together with $\rho$ and $\sigma$ and the two independent 
components of the unit vector $n^\alpha$, both sides of (\ref{dec1})
engage six field degrees of freedom. 

In order to confirm that the six field degrees of freedom 
on the {\it r.h.s.} of (\ref{dec1})
are actually independent, we first substitute the 
decomposition in the Yang-Mills field strength tensor. This
yields
\ba
F^\alpha_{ij} = ( G_{ij} - [1-(\rho^2 +
\sigma^2)] H_{ij}) \ n^\alpha \ \ \ \ \ \ \ \ \ \ \ \ \ \ \ \ \ \ \ 
\ \ \ \ \ \ \ \ \ \nn \\
+ \nabla_i \rho \partial_j n^\alpha + \nabla_i \sigma 
\epsilon^{\alpha\beta\gamma}\partial_j
n^\beta n^\gamma \ - \ \nabla_j \rho \partial_i
n^\alpha - \nabla_j \sigma \epsilon^{\alpha\beta\gamma}
\partial_i n^\beta n^\gamma
\la{ort}
\ea
Here
\[
G_{ij} \ = \ \partial_i C_j - \partial_j C_i \\ \nn
H_{ij} \ = \ \epsilon^{\alpha\beta\gamma} n^\alpha 
\partial_i n^\beta \partial_j n^\gamma
\]
and
\[ 
(\partial_i + i C_i) (\rho + i \sigma) \ = \ \nabla_i(\rho
+ i \sigma) \ \equiv \ 
\nabla_i \phi 
\]
We then substitute this decomposition of $F^\alpha_{ij}$
in the Yang-Mills action
\be
S \ = \ \frac{1}{4}\int d^2x (F_{ij}^\alpha)^2
\la{ymact}
\ee
When we perform a variation of this action {\it w.r.t.} the component
fields $(C_i,\phi,n^\alpha)$, the ensuing critical points
lead to a set of Euler-Lagrange equations. These equations
reproduce the full two dimensional Yang-Mills equations 
\ba
D^{\alpha\beta}_i F^\beta_{ij} \ = \ 0
\la{YMeq}
\ea
only when the decomposition (\ref{dec1}) is complete in directions 
which are orthogonal to the gauge orbits
\be
A^\alpha_i \ \to \ A^\alpha_i + D^{\alpha\beta}_i \epsilon^\beta \ \equiv \
A^\alpha_i + (\delta^{\alpha\beta} \partial_i + 
\epsilon^{\alpha\gamma\beta}A^\gamma_i)
\epsilon^\beta
\la{gaugetr}
\ee

The variation of the action (\ref{ymact}) 
{\it w.r.t.} the components $(C_i,\phi,n^a)$ gives 
the following Euler-Lagrange equations \cite{fadprl}
\ba
\bfn \cdot D_i \bfF_{ij} \ = \ 0
\nn \\
\kappa^+_j \ \vece_+ \cdot D_i \bfF_{ij} \ = \ 0
\nn \\
\nabla_j \phi \ \vece_- \cdot D_i \bfF_{ij} \ = \ 0 
\label{fuleq}
\ea
Here $(\vece_\theta, \vece_\varphi, \vecn)$ is a right-handed orthonormal
triplet and
\ba
\kappa^+_i \ = \ \kappa_i^\theta + i \kappa_i^\varphi \ = \ 
(\vece_\theta + i \vece_\varphi) \cdot \partial_i \vecn \ = \ 
\vece_+ \cdot \partial_i \vecn
\label{kappa}
\ea
Note that there is some latitude in the definition of
$\vece_+ = \vece_\theta + i \vece_\varphi$, without 
affecting any of our subsequent conclusions we can send
\be
\vece_+ \ \to \ e^{i\xi} \vece_+
\la{uint1}
\ee
Thus we have an internal U(1) gauge 
structure which has been discussed in \cite{fadde2}.

Since $F^\alpha_{12} = - F^\alpha_{21}$ we immediately find that the
only nontrivial regular solution to the equations (\ref{fuleq})
is the homogeneous one,
\[
(\ref{fuleq}) \ \ \ \Rightarrow \ \ \ D_i^{\alpha\beta}F^\beta_{ij} = 0
\]
that is, the full two dimensional Yang-Mills equation (\ref{YMeq}).
This means that the decomposition (\ref{dec1}) is indeed 
complete in the space of gauge orbits of $A^\alpha_i$. 

For a total completeness of the decomposition (\ref{dec1})
we still need to identify in it 
the SU(2) gauge orbit (\ref{gaugetr}). 
This gauge orbit involves three field degrees of freedom. 
One of these is the U(1) gauge transformation in the direction
of $n^\alpha$, with $\epsilon^\alpha = \epsilon n^\alpha$. It sends
\[
C_i \ \to \ C_i - \partial_i \epsilon
\]
\[
\phi = \rho + i \sigma \ \to \ e^{i\epsilon} \phi
\]
while $n^\alpha$ itself remains intact. Notice that as
a consequence ($C_i,\phi$) has a 
natural interpretation as an (electric) Abelian Higgs multiplet. 
Since the unit vector $n^\alpha$ has a natural magnetic 
interpretation (it appears as an order parameter 
{\it e.g.} in the Heisenberg model) the two 
sets of variables $(C_i,\phi)$ and $n^\alpha$ are inherent 
electric and magnetic dual variables in the two dimensional gauge theory.

The remaining two field degrees of freedom along the gauge
orbit must be orthogonal to $n^\alpha$. They can be described as 
follows: We introduce $g \in$ SU(2)
by
\[
n^\alpha \tau^\alpha \ = \ g \tau^3 g^{-1}
\]
which is manifestly U(1) invariant, {\it i.e.}
invariant under conjugation $g \to gh$ 
by an element $h \in SU(2)$ in the Cartan 
direction $\tau^3$. This corresponds to 
the U(1) gauge transformation along $n^\alpha$. 

We introduce the right-invariant form
\[
R_i \ = \ g^{-1} \partial_i g
\]
With $R^{diag}_i$ the diagonal part of $R_i$ and $R^{off}_i$
its off-diagonal part, we can write the gauge field (\ref{dec1}) as
\be
A^\alpha_i\tau^\alpha 
\ = \ g \left( C_i \tau^3 + i R^{diag}_i + \rho [R_i,
\tau^3 ] - i \sigma R^{off}_i \right) g^{-1} + i g \partial_i
g^{-1}
\la{gauA}
\ee
Consequently (\ref{dec1}) is manifestly gauge-equivalent to
(in the sequel we always have $a,b = 1,2$)
\be
B_i^\alpha\tau^\alpha 
\ = \  C_i \tau^3 + i R^{diag}_i + \rho [R_i,
\tau^3 ] - i \sigma R^{off}_i \ \equiv \ W_i \tau^3 \ + 
\ Q_i^a \tau^a
\la{gauB}
\ee
This reveals that the parametrization (\ref{dec1}), (\ref{gauA}) 
is indeed complete, also on the gauge orbit space.

Clearly, the sine-Gordon decomposition (\ref{sgdec}) must be contained
in (\ref{gauA}), (\ref{gauB}). Comparing (\ref{sgdec})
with (\ref{gauB}) we conclude that we must 
choose $\rho, \sigma$ and $n^\alpha$ such 
that
\ba
k_2 \sin \frac{\omega}{2} - i k_1 \cos \frac{\omega}{2} \ = 
\ (\rho + i \sigma) (\kappa^\theta_1 + i 
\kappa^\varphi_1) \ \equiv \ \phi \kappa^+_1 \nn
\\
k_1 \sin \frac{\omega}{2} - i k_2 \cos \frac{\omega}{2} \ = 
\ (\rho + i \sigma) (\kappa^\theta_2 +
i \kappa^\varphi_2) \ \equiv \ \phi \kappa^+_2
\la{sggau}
\ea
We parametrize
\be
\vecn \ = \left( \matrix{ \cos\varphi \sin\theta \cr
\sin\varphi \sin \theta \cr
\cos \theta }\right)
\nn
\ee
and we select the phase (\ref{uint1}) so that
(\ref{kappa}) becomes 
\[
\kappa^+_i \ = \ \kappa^\theta_i + i \kappa^\varphi_i
\ = \ \partial_i \theta + i \sin \theta \partial_i \varphi
\] 
We then get from (\ref{sggau})
\[
\left( \matrix{ \partial_1 \theta \cr \sin\theta 
\partial_1\varphi} \right) \ = \ \frac{1}{\rho^2 + \sigma^2}
\left[ \matrix{ \rho & \sigma \cr -\sigma & \rho }\right]
\left( \matrix{ k_2 \sin \frac{\omega}{2} \cr k_1 \cos
\frac{\omega}{2} } \right)
\]
\[
\left( \matrix{ \partial_2 \theta \cr \sin\theta 
\partial_2\varphi} \right) \ = \ \frac{1}{\rho^2 + \sigma^2}
\left[ \matrix{ \rho & \sigma \cr -\sigma & \rho }\right]
\left( \matrix{ k_1 \sin \frac{\omega}{2} \cr k_2 \cos
\frac{\omega}{2} } \right)
\]
from which we can solve $\theta$ and $\varphi$ in terms
of $\omega$ which is a solution of the sine-Gordon equation, and
$\rho$ and $\sigma$ which we can select quite liberally. 

\vskip 1.5cm
\noindent
{\bf \Large 3. Three Dimensional Riemann Tensor}
\vskip 1.0cm

We now proceed to show that the Yang-Mills 
field strength tensor (\ref{ort}) can be viewed as
the Riemannian curvature tensor in a three dimensional
ambient space in the immediate vicinity of a two dimensional
hypersurface. For a suggestive correspondence, 
we start by interpreting the Yang-Mills field $A^\alpha_i$ 
as a linear combination of a background field
$W_i$ which is the Cartan component in (\ref{gauB}),
and a fluctuation field $Q^a_i$ which corresponds to the off-diagonal
part in (\ref{gauB}).
The Yang-Mills field strength tensor then decomposes 
into the following Cartan part and off-diagonal part
\be
F^3_{ij} \ = \ \partial_i W_j - \partial_j W_i 
+ (Q^1_i Q^2_j - Q^2_j Q^1_i) \ \equiv \ F_{ij} + 
(Q^1_i Q^2_j - Q^2_j Q^1_i)
\la{esm1}
\ee
\be
\hskip 3.5cm F^a_{ij} \ = \ ({\delta^a}_b \partial_i - 
W_i {\epsilon^a}_b ) Q^b_j - ({\delta^a}_b \partial_j
- W_j {\epsilon^a}_b )Q^b_i \ \ \ \ \ \ \ \ (a,b=1,2)
\la{esm2}
\ee
It is instructive to compare this with (\ref{ort}), which
is the repsesentation of (\ref{esm1}), (\ref{esm2}) in
the gauge $\tau^3 \to n^\alpha \tau^\alpha$. The structural
similarity is evident.

Next, we recall Ricci's identity which states
\be
(\nabla_\rho \nabla_\sigma - \nabla_\sigma \nabla_\rho)
\partial_\eta x^\kappa \ = \ \partial_\tau 
x^\kappa {R^\tau}_{\eta\rho\sigma}
\la{ricci}
\ee
for a connection $\nabla_\rho$ 
and the ensuing curvature tensor $ {R^\tau}_{\eta\rho\sigma}$.
We employ this in the Gauss equation
(\ref{gwe}), by selecting for $\nabla_\rho$ the induced 
covariant derivative on the hypersurface. This gives
for the Riemann curvature tensor of the ambient space
the decomposition
\be
{\hat R^\mu}_{\nu\rho\sigma} \partial_i x^\nu
\partial_j x^\rho \partial_k x^\sigma
\ = \ \partial_l x^\mu {U^l}_{ijk}
+ N^\mu V_{ijk}
\la{riem1}
\ee
where
\ba
U_{lijk} \ = \ R_{lijk}
\ + \ (Q_{ij}Q_{kl} - Q_{ik}
Q_{jl}) 
\la{defU} \\
V_{ijk} \ = \ ( \delta^l_i 
\partial_k - \Gamma^l_{ik}) {Q}_{lj}
\ - \ (\delta^l_i \partial_j - 
\Gamma^l_{ij}){Q}_{lk} 
\la{defV}
\ea
with $R_{lijk}$ the 
Riemann tensor on the two dimensional surface 
and $Q_{ij}$ its extrinsic curvature.

Clearly, there is a definite formal similarity 
between (\ref{esm1}) and (\ref{defU}), and between
(\ref{esm2}) and (\ref{defV}) suggesting that 
we can relate $F^3_{ij} \sim U_{lijk}$ 
and $F^a_{ij} \sim V_{ijk}$. 
If this identification indeed holds, 
the two-dimensional Yang-Mills field strength tensor 
computes the three dimensional ambient Riemann curvature tensor:
The ${U}_{lijk}$ is the restriction 
of the Riemann tensor to the 
tangent of the surface, and $V_{ijk}$ is the 
projection of the Riemann tensor along the unit normal 
of the surface. Consequently we obtain
the entire three dimensional Riemann curvature tensor 
from the two dimensional $F^a_{ij}$, 
in the vicinity of the two dimensional hypersurface.
In this sense the three dimensional 
gravity is then induced by the two dimensional Yang-Mills theory.

We shall now proceed to establish the relations between
(\ref{esm1}), (\ref{esm2}) and (\ref{defU}), (\ref{defV}).
For this we denote by $u,v,...  = 1,2$ 
a local frame (tangent bundle) on the two dimensional 
hypersurface in the three dimensional ambient space. 
The ensuing zweibein obeys ${e^u}_i {e^v}_j 
\eta_{uv} \ = \ g_{ij}$ and ${e^u}_i {E^i}_v  \ 
= \ {\delta^u}_v$ {\it etc}. 
Furthermore, we introduce ${\epsilon^u}_v$ with ${\epsilon^1}_2 = 
- {\epsilon^2}_1 = 1$. We also introduce the zweibein
${e^a}_u$ with inverse $ {E^u}_a$, which relate the local frame
of the hypersurface to the
off-diagonal part of the SU(2) Lie-algebra.

We start with the decomposition (\ref{gauB}), where we write
\[
Q^{1,2}_i \ \equiv \ {Q^a}_i \ = \ {e^a}_u {Q^u}_i \ = \
{e^a}_u {e^u}_j {Q^j}_i
\]
We then consider $F^a_{ij}$. From (\ref{esm2}) we get
\ba
{E^v}_a F^a_{ij} \ = \ \partial_i {Q^v}_j + 
({E^v}_a\partial_i {e^a}_u  - W_i {\epsilon^v}_u){Q^u}_j
\ - \ (i\leftrightarrow j)
\nn
\ea
\ba
= \ \eta^{vw} {E^k}_w \partial_i Q_{kj} \ + \
\left( [ {E^v}_a \partial_i {e^a}_u - W_i {\epsilon^v}_u ]
\eta^{uw} {E^k}_w + \eta^{vw} \partial_i {E^k}_w
\right) Q_{k j} \ - \ ( i 
\leftrightarrow j )
\nn
\ea

Consider 
\[
{E^v}_a \partial_i {e^a}_u - W_i {\epsilon^v}_u 
\]
Here ${e^a}_u$ is a zweibein between two-dimensional flat
Euclidean spaces, and it can be represented explicitely 
{\it e.g.} as
\[
{e^1}_u \ = \ \left( \matrix{ \cos \psi \cr \sin \psi } \right)
\ \ \ \ \ \& \ \ \ \ \ 
{e^2}_u \ = \ \left( \matrix{ -\sin \psi \cr \cos \psi } \right)
\]
so that 
\[
{E^v}_a \partial_i {e^a}_u - W_i {\epsilon^v}_u \ = \ 
- (W_i - \partial_i \psi) {\epsilon^v}_u
\]
This suggests that we introduce a U(1) gauge transformation
in the Cartan direction of SU(2), and redefine
\be
W_i \ - \partial_i \psi \ \to \ W_i
\la{u1}
\ee
This gives
\be
{e^u}_k \eta_{uw} {E^w}_a F^a_{ij}
\ = \ \partial_i Q_{kj} - ( {E^l}_u \partial_i
{e^u}_k - W_i {\epsilon^l}_{k} 
) Q_{lj} \ - \  ( i
\leftrightarrow j )
\la{step2}
\ee
We recall the familiar relation between spin 
connection and Christoffel symbol,
\[
\Gamma^l_{ki} \ = \ {\omega^l}_{ki}
\ + \ {E^l}_u\partial_i {e^u}_k
\]
Hence, if we identify
\be
{\omega^l}_{ki} \ = \ - W_i {\epsilon^l}_k
\la{spin}
\ee
we can write (\ref{step2}) as
\be
{e^u}_k \eta_{uv} {E^v}_a F^a_{ij} \ = \ 
[\partial_i Q_{kj} - \Gamma^l_{ik} Q_{lj}]
\ - \ [\partial_j Q_{ki} - \Gamma^l_{jk} Q_{li}]
\la{offriem}
\ee
Thus 
\be
F^a_{ij} \ = \ {e^a}_u \eta^{uv} {E^k}_v V_{jik}
\la{FV}
\ee
and consequently the off-diagonal part of the two-dimensional
Yang-Mills field strength tensor 
computes the tangential part (\ref{defU})
of the three dimensional Riemann curvature tensor (\ref{riem1}).

We now proceed to inspect the Cartan component $F^3_{ij}$ of
the Yang-Mills field strength tensor. When we recall
the representation of the Riemann tensor in terms of the 
spin connection, we get
\ba
{R^u}_{vij} \ = \ \partial_i {\omega^u}_{vj}
- \partial_j {\omega^u}_{vi} + {\omega^u}_{wi}
{\omega^w}_{vj} - {\omega^u}_{wj} 
{\omega^w}_{vi} \ = \ - (\partial_i W_j - 
\partial_j W_i) {\epsilon^u}_v
\nn
\ea
where we have used (\ref{spin}). But from (\ref{esm1})
we now get immediately the desired relation
\be
F^3_{ij} \ = \ \frac{1}{2} {\epsilon^v}_u ( {R^u}_{vij}
- {Q^u}_i Q_{vj} + {Q^u}_j Q_{vi} )
\ = \ \frac{1}{2} {\epsilon^v}_u {U^u}_{vij}
\la{FU}
\ee
and we conclude that $F^3_{ij}$ indeed computes the normal
component of the three dimensional Riemann tensor.

When we combine (\ref{FU}) with (\ref{FV}) we arrive at our
main result: The two dimensional 
Yang-Mills field strength tensor can be interpreted as
a three dimensional Riemann curvature tensor in the vicinity
of the two dimensional hypersurface. In this 
sense, the two dimensional Yang-Mills theory is then
a source of gravity in the three dimensional ambient space.

\vskip 1.5cm
\noindent
{\bf \Large 4. Further Developments}
\vskip 1.0cm
The present results can be extended in a variety of directions.
For example, the flatness condition (\ref{zeroF}) in the 
two-dimensional gauge theory can also 
be interpreted as the equation of 
motion (first class constraint) in the three dimensional 
SU(2) Chern-Simons theory, when wiewed 
as a Hamiltonian system 
\be
S \ = \ \int d^3x Tr[ A \wedge dA + \frac{2}{3} A^3 ]
\ \to \ \int d^2xdt \left(
\epsilon^{ij} A_i^\alpha \partial_t A^\alpha_j \ - \
A^\alpha_0 \epsilon^{ij}F^\alpha_{ij} \right) 
\la{chs}
\ee
Here we have shown that the condition (\ref{zeroF}) 
can also be identified with the Gauss-Codazzi equations, 
which modulo rigid rotations and translations describe 
the embedding of two dimensional Riemann surfaces
in flat three dimensional $R^3$. Consequently
the Chern-Simons theory determines the Hamiltonian dynamics 
of two dimensional Riemann surfaces in flat $R^3$.
Since the condition (\ref{zeroF}) also relates to the Lax
pair of integrable models,
the dynamics of these Riemann surfaces is integrable, and
the surfaces scatter from each other in an elastic manner which
directly relates to the properties of conventional two 
dimensional integrable models \cite{fadbook}. 
The SU(2) Chern-Simons theory also describes three 
dimensional knot invariants \cite{witcs} suggesting 
interesting connections between knot theory and
the dynamics of two dimensional Riemann surfaces in $R^3$.

Finally, since (\ref{zeroF}) describes the embedding
of Riemann surfaces in a flat three dimensional space the
ensuing Chern-Simons theory does not employ four dimensional
gravity. It would be very interesting to develop a generalization
of the Chern-Simons theory, with four dimensional gravity included.
This generalization should lead to (\ref{riem1}), 
(\ref{FV}), (\ref{FU}) as its equations of motion, describing 
the dynamics of two dimensional Riemann surfaces in four 
dimensional curved ambient space with a 
curvature induced by the Riemann surfaces, radiating gravity.

\vskip 1.2cm
\noindent
{\bf \Large 5. Conclusions}
\vskip 1.0cm
In conclusion, we have shown that the two dimensional SU(2)
Yang-Mills field strength tensor can be interpreted as a
three dimensional Riemann curvature tensor. This can be
further interpreted so that the two dimensional gauge theory
is a source of three dimensional gravity. A vanishing
Yang-Mills field strength tensor then leads to a vanishing Riemannian
curvature, and consequently it has
an interpretation in terms of the original
Gauss-Codazzi equations which describe the isometric embedding of
Riemann surfaces in flat $R^3$. The vanishing Yang-Mills field
strength tensor also yields 
a SU(2) Lax pair which implies that two dimensional integrable models
with such a Lax pair specify Riemann surfaces in flat
$R^3$. Furthermore,
since a vanishing two dimensional field strength tensor also arises as
the Hamiltonian equation of motion in three dimensional Chern-Simons
theory, this theory admits an interpretation
in terms of Hamiltonian dynamics of two dimensional 
Riemann surfaces in flat four dimensional ambient space. Obviously it 
would be interesting to generalize the Chern-Simons theory so
that it allows for a nontrivial four dimensional curvature.    

\vskip 1.5cm
We thank J.M. Maillet and K. Zarembo for discussions and L. Faddeev
for comments. This work was
completed while the author visited Ecole Normale Superieure in Lyon,
and we thank M. Magro for hospitality.

\vskip 2.0cm

\vfill\eject

\vfill\eject

}

\end{document}